\documentclass{aa}
\usepackage{graphicx}
\usepackage{natbib}
\usepackage{bm,times,color}
\graphicspath{{./fig/}{./png/}}

\newcommand{\EQ}{\begin{equation}}
\newcommand{\EN}{\end{equation}}
\newcommand{\EQA}{\begin{eqnarray}}
\newcommand{\ENA}{\end{eqnarray}}

\newcommand{\Eq}[1]{Equation~(\ref{#1})}

\newcommand{\Fig}[1]{Figure~\ref{#1}}

\newcommand{\Tab}[1]{Table~\ref{#1}}

\newcommand{\meanrho}{\overline{\rho}}

{}
{}
{}

\newcommand{\meanSSSS}{\overline{\mbox{\boldmath ${\mathsf S}$}} {}}

\newcommand{\meanSSS}{\overline{\mathsf{S}}}
{}
{}
{}
{}
{}
{}
{}
{}
\newcommand{\meanAA}{\overline{\mbox{\boldmath $A$}}{}}{}
\newcommand{\meanBB}{\overline{\mbox{\boldmath $B$}}{}}{}
{}
{}
{}
{}
{}
{}
{}
{}
\newcommand{\meanJJ}{\overline{\mbox{\boldmath $J$}}{}}{}
{}
\newcommand{\meanUU}{\overline{\mbox{\boldmath $U$}}{}}{}

\newcommand{\meanWW}{\overline{\mbox{\boldmath $W$}}{}}{}
{}
{}
\newcommand{\meanA}{\overline{A}}
\newcommand{\meanB}{\overline{B}}

\newcommand{\meanh}{\overline{h}}

\newcommand{\meanH}{\overline{H}}
\newcommand{\meanU}{\overline{U}}

\newcommand{\meanT}{\overline{T}}

{}

{}
{}

\newcommand{\pphi}{\hat{\vec{\phi}}}

\newcommand{\nab}{\mbox{\boldmath $\nabla$} {}}

\newcommand{\meanQQQ}{\overline{\mbox{\boldmath ${\cal Q}$}} {}}
\newcommand{\DD}{{\rm D} {}}

\newcommand{\dd}{{\rm d} {}}

\newcommand{\const}{{\rm const}  {}}

\def\degr{\hbox{$^\circ$}}

\def\csz{c_{\rm s0}}
\def\Hp{H_{\rm p}}
\def\Hpz{H_{\rm p0}}
\def\cp{c_{\rm p}}

\def\qpz{q_{\rm p0}}
\def\qp{q_{\rm p}}

\def\rstar{r_\star}

\def\betap{\beta_{\rm p}}

\def\betastar{\beta_{\star}}

\def\peff{p_{\rm eff}}

\def\kf{k_{\rm f}}

\def\urms{u_{\rm rms}}

\def\nut{\nu_{\rm t}}
\def\nuT{\nu_{\rm T}}

\def\etat{\eta_{\rm t}}

\def\etaT{\eta_{\rm T}}

\def\Beq{B_{\rm eq}}
\def\Beqz{B_{\rm eq0}}

\def\half{{\textstyle{1\over2}}}

\def\onethird{{\textstyle{1\over3}}}

\newcommand{\Mm}{\,{\rm Mm}}

\newcommand{\etal}{et al.}
\newcommand{\yapj}[3]{ #1, {ApJ,} {#2}, #3}

\newcommand{\yapjl}[3]{ #1, {ApJ,} {#2}, #3}

\newcommand{\yan}[3]{ #1, {Astron.\ Nachr.,} {#2}, #3}

\newcommand{\yana}[3]{ #1, {A\&A,} {#2}, #3}

\newcommand{\yanar}[3]{ #1, {A\&A Rev.,} {#2}, #3}

\newcommand{\ygafd}[3]{ #1, {Geophys.\ Astrophys.\ Fluid Dyn.,} {#2}, #3}

\newcommand{\ypf}[3]{ #1, {Phys.\ Fluids,} {#2}, #3}

\newcommand{\ysovl}[3]{ #1, {Sov.\ Astron.\ Lett.,} {#2}, #3}
\newcommand{\yjetp}[3]{ #1, {Sov.\ Phys.\ JETP,} {#2}, #3}

\newcommand{\yptrs}[3]{ #1, {Phil.\ Trans.\ R.\ Soc.,} {#2}, #3}

\newcommand{\ymn}[3]{ #1, {MNRAS,} {#2}, #3}
\newcommand{\ynat}[3]{ #1, {Nature,} {#2}, #3}

\newcommand{\ysph}[3]{ #1, {Solar Phys.,} {#2}, #3}

\newcommand{\ypre}[3]{ #1, {Phys.\ Rev.\ E,} {#2}, #3}

\newcommand{\yjour}[4]{ #1, {#2}, {#3}, #4}

\newcommand{\ybook}[3]{ #1, {#2} (#3)}

\titlerunning{}
\authorrunning{S. Jabbari \etal}
\title{Surface flux concentrations in a spherical $\alpha^2$ dynamo}
\author{S. Jabbari\inst{1,2}, A. Brandenburg\inst{1,2},
N. Kleeorin\inst{1,3,4}, D. Mitra\inst{1}, and I. Rogachevskii\inst{1,3,4}
}
\institute{
Nordita, KTH Royal Institute of Technology and Stockholm University,
Roslagstullsbacken 23, 10691 Stockholm, Sweden
\and
Department of Astronomy, AlbaNova University Center,
Stockholm University, 10691 Stockholm, Sweden
\and
Department of Mechanical Engineering, Ben-Gurion University of the Negev,
POB 653, Beer-Sheva 84105, Israel
\and
Department of Radio Physics, N.~I.~Lobachevsky State University of
Nizhny Novgorod, Russia
}
\date{\today,~ $ $Revision: 1.1 $ $}
\begin{document}

\abstract{
In the presence of strong density stratification, turbulence can lead
to the large-scale instability of a horizontal magnetic field if its
strength is in a suitable range
(around a few percent of the turbulent equipartition value).
This instability is related to a suppression of the turbulent pressure
so that the turbulent contribution
to the mean magnetic pressure
becomes negative.
This results in the excitation of a
negative effective magnetic pressure instability (NEMPI).
This instability has so far only been studied for an imposed magnetic field.
}{
We want to know how NEMPI works when the mean magnetic
field is generated self-consistently by an $\alpha^2$
dynamo, whether it is affected by global spherical geometry,
and whether it can influence the properties of the dynamo itself.
}{
We adopt the mean-field approach, which has previously been shown to
provide a realistic description of NEMPI in direct numerical simulations.
We assume axisymmetry and solve the mean-field equations with the
{{\sc Pencil Code}} for an adiabatic stratification at a total density
contrast in the radial direction of $\approx4$ orders of magnitude.
}{
NEMPI is found to work when the dynamo-generated field is about 4\%
of the equipartition value, which is achieved through strong $\alpha$
quenching.
This instability is excited
in the top 5\% of the outer radius, provided the density
contrast across this top layer is at least 10.
NEMPI is found to occur at lower latitudes when the
mean magnetic field is stronger.
For weaker fields, NEMPI can make the dynamo oscillatory with poleward
migration.
}{
NEMPI is a viable mechanism for producing magnetic flux concentrations
in a strongly stratified spherical shell in which a magnetic field is
generated by a strongly quenched $\alpha$ effect dynamo.
}

\maketitle

\section{Introduction}

The magnetic field of stars with outer convection
zones, including that of the Sun, is believed to
be generated by differential rotation and
cyclonic convection \citep[see, e.g.,][]{Mof78,Par79,ZRS83,BS05}.
The latter leads to an $\alpha$ effect, which refers to an important new
term in the averaged (mean-field) induction
equation, quantifying the component of the mean
electromotive force that is aligned with the mean
magnetic field \citep[see, e.g.,][]{SKR66,KR80,BGR13}.
However, what is actually
observed are sunspots and active regions, and
the description of these phenomena is not part of
conventional mean-field dynamo theory
\citep[see, e.g.,][]{PR82,S89,O03,Cally03,SK12}.

Flux tube models \citep{P55,P82,P84,SW80,S81,SCM94,DC99}
have been used to explain the formation of active regions and sunspots
in an ad hoc manner.
It is then simply assumed that a sunspot emerges when the magnetic field
of the dynamo exceeds a certain threshold just above the bottom of the
convection zone for the duration of about a month \citep{CNC04}.
Such models assume the existence of strong
magnetic flux tubes at the base of the convection zone.
They require magnetic fields with a strength of
about $10^5$ gauss \citep{DSC93}.
However, such strong magnetic fields are highly unstable \citep{ASR05}
and are also difficult to produce by
dynamo action in turbulent convection \citep{GK11}.

Another possible mechanism for producing
magnetic flux concentrations is
the negative effective magnetic pressure
instability (NEMPI), which can occur in the
presence of strong density stratification, i.e.,
usually near the stellar surface, on
scales encompassing those of many turbulent
eddies.
NEMPI is caused by the suppression of turbulent
magnetohydrodynamic pressure (the isotropic part of combined Reynolds
and Maxwell stresses) by the mean magnetic field.
At large Reynolds numbers, the negative turbulent contribution
can become so large that the effective mean magnetic pressure
(the sum of turbulent and nonturbulent contributions) is negative.
This results in the excitation of NEMPI
that causes formation of large-scale inhomogeneous magnetic
structures.
The instability mechanism is as follows.
A rising magnetic flux tube expands, the field becomes weaker,
but because of negative magnetic pressure, its magnetic pressure increases,
so the density decreases, and it becomes lighter still and rises further.
Conversely, a sinking tube contracts, the magnetic field increases,
but the magnetic pressure decreases, so the density increases,
and it becomes heavier and sinks further.
The energy for this instability is supplied by the small-scale turbulence.
By contrast, the free energy in Parker's magnetic buoyancy instability
or in the interchange instability in plasma,
is drawn from the gravitational field \citep{N61,P66}.

Direct numerical simulations \citep[DNS;
see][]{BKKMR11,KBKMR12b}, mean-field simulations
\citep[MFS; see][]{BKR10,BKKR12,KBKMR12c,Kapy12}, and
earlier analytic studies \citep{KRR89,KRR90,KMR96,KR94,RK07}
now provide conclusive evidence for the physical reality of NEMPI.
However, open questions still need to be
answered before it can be applied to detailed
models of active regions and sunspot formation.

In the present paper we take a first step toward combining NEMPI,
which is described well using mean-field theory,
with the $\alpha$ effect in mean-field dynamos.
To study the dependence of NEMPI on the magnetic field strength,
we assume that $\alpha$ is quenched.
This allows us to change the magnetic field strength by changing
the quenching parameter.
We employ spherical coordinates $(r,\theta,\phi)$, with radius $r$,
colatitude $\theta$, and azimuthal angle $\phi$.
We assume axisymmetry, i.e., $\partial/\partial\phi=0$.
Furthermore, $\alpha$ is a pseudo-scalar that changes sign
at the equator, so we assume that $\alpha$ is proportional to
$\cos\theta$, where $\theta$ is the colatitude \citep{Rob72}.
We arrange the quenching of $\alpha$ such that the resulting mean
magnetic field is in the appropriate interval to allow NEMPI to work.
This means that the effective (mean-field) magnetic pressure locally has a negative derivative with respect to increasing normalized
field strength \citep{KBKMR12c}, so the mean toroidal
magnetic field must be less than about 20\% of the equipartition
field strength.

The choice of using spherical geometry is taken because the
dynamo-generated magnetic field depends critically on the geometry.
Therefore, to have a more realistic field structure, we felt it profitable
to carry out our investigations in spherical geometry.
Guided by the insights obtained from such studies, it will in future
be easier to design simpler Cartesian models to address specific questions
regarding the interaction between NEMPI and the dynamo instability.

In the calculations presented below we use the
{\sc Pencil Code}\footnote{http://pencil-code.googlecode.com}, which has been used in
DNS of magneto-hydrodynamics in spherical coordinates~\citep{MTBM09} and also
in earlier DNS and MFS of NEMPI.
Unlike most of the earlier calculations, we adopt an adiabatic
equation of state.
This results in a stratification such that the temperature declines
approximately linearly toward the surface, so the scale height becomes
shorter and the stratification stronger toward the top layers.
This is done to have a clear segregation between the dynamo in the bulk
and NEMPI near the surface, where the stratification is strong enough
for NEMPI to operate.
The gravitational potential is that of a point mass.
This is justified because the mass in the convection zone is
negligible compared to the one below.
The goal of the present work is to produce reference cases in spherical
geometry and to look for
new effects of spherical geometry.
We begin by describing the basic model.

\section{The model}

The evolution equations for mean vector potential $\meanAA$,
mean velocity $\meanUU$, and mean density $\meanrho$, are
\EQA
\label{dAmean}
{\partial\meanAA\over\partial t}&=&\meanUU\times\meanBB
+\alpha\meanBB-\etaT\meanJJ,\\
\label{dUmean}
{\DD\meanUU\over\DD t}&=&{1\over\meanrho}\left[
\meanJJ\times\meanBB+\nab(q_{\rm p}\meanBB^2/2\mu_0)\right]
-\nuT\meanQQQ-\nab\meanH,\\
{\DD\meanrho\over\DD t}&=&-\meanrho\nab\cdot\meanUU,
\ENA
where $\DD/\DD t=\partial/\partial t+\meanUU\cdot\nab$
is the advective derivative, $\meanrho$ is the mean density,
$\meanH=\meanh+\Phi$ is the mean reduced enthalpy
with $\meanh=c_p\meanT$  the mean enthalpy,
$\meanT\propto\meanrho^{\gamma-1}$  the mean temperature,
$\gamma=c_p/c_v$ is the ratio of specific heats
at constant pressure and constant density, respectively,
$\Phi$ is the gravitational potential,
$\etaT=\etat+\eta$ and $\nuT=\nut+\nu$ are the sums of turbulent and
microphysical values of magnetic diffusivity and kinematic viscosities,
respectively, $\alpha$ is the aforementioned coefficient
in the $\alpha$ effect,
$\meanJJ=\nab\times\meanBB/\mu_0$  is the mean current density,
$\mu_0$ is the vacuum permeability,
\EQ
-\meanQQQ=\nabla^2\meanUU+\onethird\nab\nab\cdot\meanUU
+2\meanSSSS\nab\ln\meanrho
\EN
is a term appearing in the viscous force, where
$\meanSSSS$ is the traceless rate of strain tensor of the mean flow
with components $\meanSSS_{ij}=\half(\meanU_{i,j}+\meanU_{j,i})
-\onethird\delta_{ij}\nab\cdot\meanUU$,
and finally $\nab(q_{\rm p}\meanBB^2/2\mu_0)$
determines the turbulent contribution to the mean Lorentz force.
Here, $q_{\rm p}$ depends on the local field strength (see below).
This term enters with a plus sign, so positive values of $\qp$
correspond to a suppression of the total turbulent pressure.
The net effect of the mean field leads to an effective
mean magnetic pressure $\peff=(1-\qp)\meanBB^2/2\mu_0$,
which becomes negative for $\qp>1$,
which can indeed be the case for magnetic Reynolds numbers well
above unity \citep{BKKR12}.

Following \cite{KBKR12}, the function $\qp(\beta)$ is approximated by
\EQ
\qp(\beta)={\qpz\over1+\beta^2/\betap^2}
={\betastar^2\over\betap^2+\beta^2},
\label{qp-apr}
\EN
where $\qpz$, $\betap$, and $\betastar=\betap\qpz^{1/2}$ are constants,
$\beta=|\meanBB|/\Beq$ is the modulus of the normalized mean
magnetic field, and $\Beq=\sqrt{\mu_0\rho}\, \urms$
is the equipartition field strength.

NEMPI can occur at a depth where the derivative, $\dd\peff/\dd\beta^2$,
is negative.
Since the spatial variation of $\beta$ is caused mainly by the
increase in density with depth, the value of the mean horizontal magnetic
field essentially determines the location where NEMPI can occur.
Therefore, the field strength has to be in a suitable range such that
NEMPI occurs within the computational domain.
Unlike the Cartesian cases investigated in earlier work
\citep{BKR10,BKKR12,KBKR12}, where it is
straightforward to impose a magnetic field, in a
sphere it is easier to generate a magnetic field
by a mean-field dynamo.
This is why we
include a term of the form $\alpha\meanBB$ in the
expression for the mean electromotive force
[second term on the righthand side of \Eq{dAmean}].
When the mean magnetic field is generated by a
dynamo, the resulting magnetic field strength
depends on the nonlinear suppression of the
dynamo. We assume here a simple quenching
function for the $\alpha$ effect, i.e.,
\begin{equation}
\alpha(\theta,\beta)={\alpha_0\cos\theta \over 1+Q_\alpha\beta^2},
\label{alpha}
\end{equation}
where $Q_\alpha$ is a quenching parameter that determines
the typical field strength,
which is expected to be on the order of $Q_\alpha^{-1/2}\Beq$.
The value of $Q_\alpha$ must be chosen large enough so that the
nonlinear equilibration of the dynamo process results in a situation such that
$\dd\peff/\dd\meanB$ is indeed negative within the computational domain.
In analogy with the $\betap$ parameter in \Eq{qp-apr}, we can define
a parameter $\beta_\alpha=Q_\alpha^{-1/2}$, which will be quoted
occasionally.

The strength of the dynamo is also determined by the dynamo number,
\EQ
C_\alpha=\alpha_0 R/\etaT.
\EN
For our geometry with $0.7\leq r/R\leq1$, the critical value of
$C_\alpha$ for the onset of dynamo action is around 18.
The excitation conditions for dipolar and quadrupolar parities
are fairly close together.
This is because the magnetic field is strongest at high latitudes,
so the hemispheric coupling is weak.
In the following we restrict ourselves to solutions with dipolar parity.
We adopt the value $C_\alpha=30$, so the dynamo is nearly twice supercritical.

\begin{figure}[t!]\begin{center}
\includegraphics[width=\columnwidth]{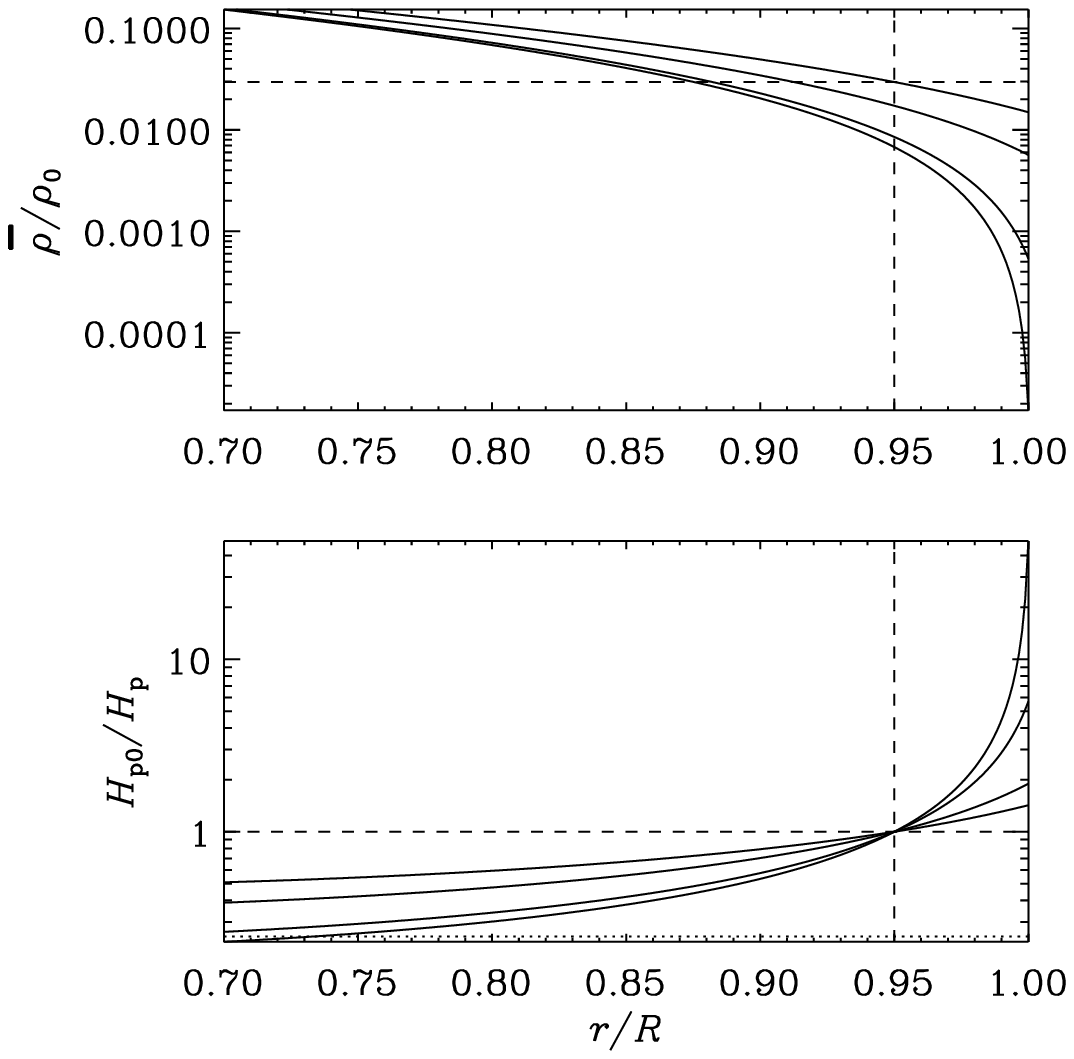}
\end{center}\caption[]{
Initial stratification of density and inverse scale height
for $\rstar/R=1.001$ (strongest stratification), 1.01, 1.05, and 1.1.
The dashed lines mark the position of the reference radius
$r_{\rm ref}=0.95\,R$,
where $\rho/\rho_0\approx0.0068$ for $\rstar/R=1.001$
and $\Hp(r)=\Hpz$ by definition.
The dotted line marks the value of $\etat/\betastar\urms\Hpz$.
}\label{pstrat}\end{figure}

As mentioned before, our gravitational potential $\Phi$ is that of a
point mass.
We define $\Phi$ such that it vanishes at a radius $\rstar$, i.e.
\EQ
\Phi(r)=-GM\left({1\over r}-{1\over\rstar}\right),
\label{Phir}
\EN
where $G$ is Newton's constant and $M$ is the mass of the sphere.
The radial component of the gravitational acceleration is then $g=-GM/r^2$.
We adopt an initially adiabatic stratification with $\cp\meanT=-\Phi(r)$,
so $\meanT$ vanishes at $r=\rstar$.
To avoid singularities, the value of $\rstar$ has to be chosen
some distance above $r=R$.
The radius $\rstar$ is used to set the density contrast.
\Tab{Tab2} gives the density contrast for different values of $\rstar$.
We vary $\rstar$ between $1.001\,R$, which corresponds to
our reference model with a density contrast of $8900$,
and $1.1\,R$, where the density contrast is 14.
The pressure scale height is given by
\EQ
\Hp(r)={r(1-r/\rstar) \over n+1},
\EN
where $n=1/(\gamma-1)=3/2$ is the polytropic index for an
adiabatic stratification with $\gamma=5/3$.
The density scale height is $H_\rho=r(1-r/\rstar)/n$.
The initial density profile is given by
\EQ
\meanrho/\rho_0=(-\Phi/n\csz^2)^n.
\EN
Radial profiles of $\meanrho/\rho_0$ and the inverse pressure
scale height $\Hpz/\Hp(r)$, are shown in \Fig{pstrat} for
$\rstar/R$ varying between 1.1 and 1.001.
Here, $\Hpz=\Hp(r_{\rm ref})$ is the pressure scale height
at the reference radius $r_{\rm ref}=0.95\,R$, corresponding to a
depth of $35\Mm$ in the Sun.

\begin{table}[b!]\caption{
Dependence of the density contrast on the value of $\rstar$.
}\vspace{12pt}\centerline{\begin{tabular}{rcccccc}
$\rstar/R$ &
$\Hp({\rm top})/R$ &
$\Hpz/R$ &
$\rho_{\max}/\rho_{\min}$ \\
\hline
$ 1.100$&$ 3.6\times10^{-2}$&$0.052$&$ 1.4\times10^{1}$\\
$ 1.010$&$ 4.0\times10^{-3}$&$0.023$&$ 2.9\times10^{2}$\\
$ 1.001$&$ 4.0\times10^{-4}$&$0.019$&$ 8.9\times10^{3}$\\
\label{Tab2}\end{tabular}}\end{table}

The analytic estimate of the growth rate of NEMPI, $\lambda$, based on
an isothermal layer with $\Hp=H_\rho=\const$ is given by \citep{KBKMR12c}
\EQ
\lambda\approx\betastar{\urms\over\Hp}-\etat k^2.
\EN
Assume that this equation also applies to the current case where
$\Hp$ depends on $r$, and setting $k=\Hpz^{-1}$,
the normalized growth rate is
\EQ
{\lambda\Hpz\over\betastar\urms}
={\Hpz\over\Hp}-{\etat\over\betastar\urms\Hpz}.
\EN
In \Fig{pstrat} we compare therefore $\Hpz/\Hp$ with
$\etat/\betastar\urms\Hpz$ and see that the former exceeds
the latter in our reference model with $\rstar/R=1.001$.
This suggests that NEMPI should be excited in the outer layers.

As nondimensional measures of $\etat$ and $\urms$, we define
\EQ
\tilde\etat=\etat/\sqrt{GMR},\quad
\tilde\urms=\urms/\sqrt{GM/R},
\EN
for which we take the values $\tilde\etat=2\times10^{-4}$
and $\tilde\urms=0.07$, respectively.
Using the estimate $\etat=\urms/3\kf$ \citep{SBS08}, our choice of $\etat$
implies that the normalized wavenumber of the energy-carrying eddies
is $\kf R=\tilde\urms/3\tilde\etat\approx120$ and that $\kf\Hpz$ varies
between 6.2 (for $\rstar/R=1.1$) and 2.3 (for $\rstar/R=1.001$).

For the magnetic field, we adopt perfect conductor boundary conditions
on the inner and outer radii, $r_0=0.7\,R$ and $R$, respectively, i.e.,
\EQ
{\partial\meanA_r\over\partial r}=\meanA_\theta=\meanA_\phi=0,
\quad\mbox{on $r=r_0,R$}.
\EN
On the pole and the equator, we assume
\EQ
{\partial\meanA_r\over\partial\theta}
=\meanA_\theta={\partial\meanA_\phi\over\partial\theta}=0,
\quad\mbox{on $\theta=0\degr$ and $90\degr$}.
\EN
Since our simulations are axisymmetric, the magnetic field is
conveniently represented via $\meanB_\phi$ and $\meanA_\phi$.
In particular, contours of $r\sin\theta\meanA_\phi$ give the
magnetic field lines of the poloidal magnetic field,
$\meanBB_{\rm pol}=\nab\times(\meanA_\phi\pphi)$.

In all cases presented in this paper, we adopt a numerical resolution
of $256\times1024$ mesh points in the $r$ and $\theta$ directions.
This is significantly higher than what has been used previously,
even in mean field calculations with stratification and hydrodynamical
feedback included; see \cite{BMT92}, where a resolution of just
$41\times81$ meshpoints was used routinely.
In principle, lower resolutions are possible, but in some cases we found
certain properties of the solutions to be sensitive to the resolution.

\begin{figure}[t!]\begin{center}
\includegraphics[width=\columnwidth]{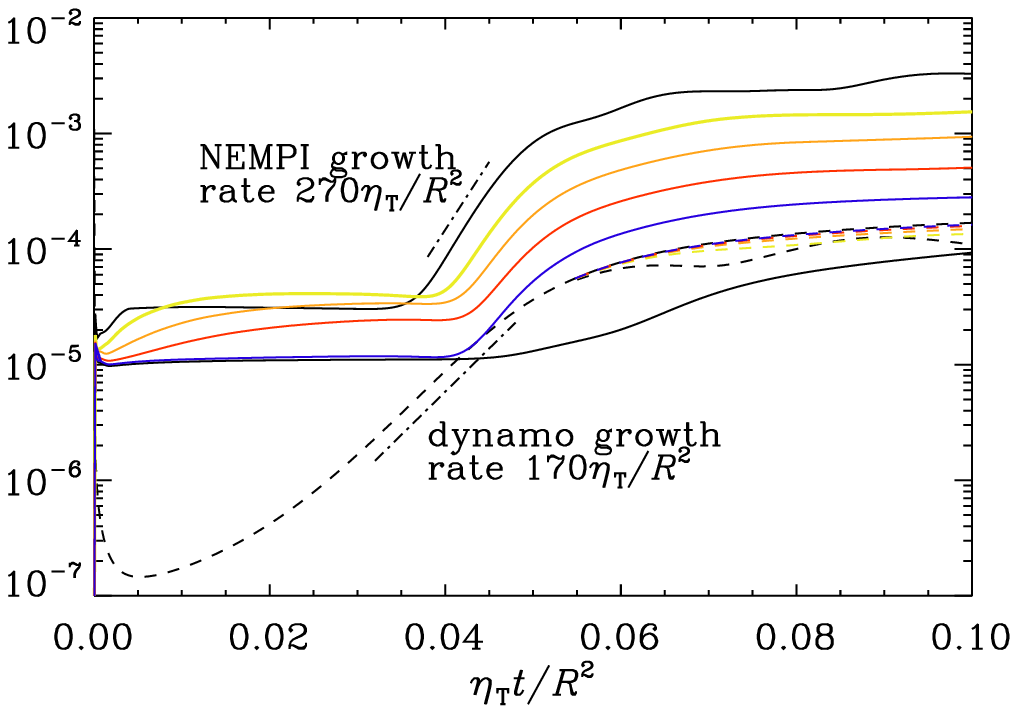}
\end{center}\caption[]{
Dependence of $\meanB_{\rm rms}$ (dashed lines) and
$\meanU_{\rm rms}$ (solid lines) on time in units of $\etaT/R^2$
for $\qpz=0$ (black), 5 (blue), 10 (red), 20 (orange), 40 (yellow),
and 100 (upper black line for $\meanB_{\rm rms}$).
The results for $\meanU_{\rm rms}$ depend only slightly on $\qpz$,
and this only when the dynamo is saturated.
}\label{pdynempi}
\end{figure}

\section{Results}

In our model, the dynamo growth rate is about $170\,\etaT/R^2$.
Although both dynamo and NEMPI are linear instabilities,
this is no longer the case in our coupled system, because
NEMPI depends on the magnetic field strength, and only in the nonlinear
regime of the dynamo does the field reach values high enough
for NEMPI to overcome turbulent magnetic diffusion.
This is shown in \Fig{pdynempi} where we plot the growth of the
magnetic field and compare with runs with different values of $\qpz$.
For $\qpz=100$ we find a growth rate of about $270\,\etaT/R^2$.
This value is significantly more than the dynamo growth rate,
and the growth occurs at the time when structures form, so
we associate this higher growth rate with that of NEMPI.

\begin{figure}[t!]\begin{center}
\includegraphics[width=\columnwidth]{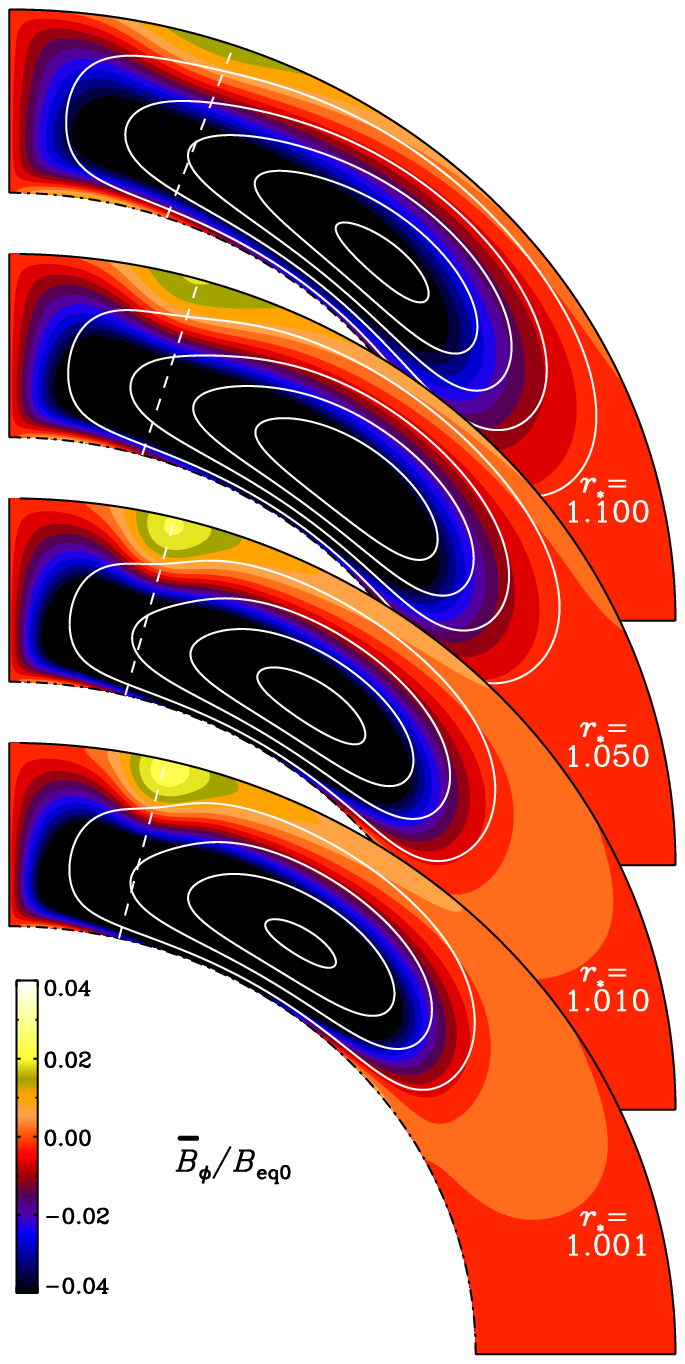}
\end{center}\caption[]{
Meridional cross-sections of $\meanB_\phi/\Beq$ (color coded)
together with magnetic field lines of $\meanBB_{\rm pol}$
for different stratification parameters $\rstar$
and $Q_\alpha=10^3$.
The dashed lines indicate the latitudes $70.3\degr$, $73.4\degr$, $75.6\degr$,
and $76.4\degr$.
}\label{pmer_bphi_256x1024P_qp100c_Q1e3_a006c_pole_str}
\end{figure}

We now discuss the resulting magnetic field structure.
We begin by discussing the effects of varying the stratification.
To see the effect of NEMPI more clearly,
we consider a somewhat optimistic set of parameters describing NEMPI,
namely $\qpz=100$ and $\betap=0.05$, which yields $\betastar=0.5$;
see \Eq{qp-apr}.
This is higher than the values 0.23 and 0.33 found from numerical simulations
with and without small-scale dynamo action, respectively \citep{BKKR12}.
The effect of lowering the value of $\qpz$ can be seen in \Fig{pdynempi}
and is also discussed below.
We choose $Q_\alpha=1000$ for the $\alpha$ quenching parameter
so that the local value of $\meanB_\phi/\Beq$ near the surface
is between 10 and 20 percent, which is suitable for exciting NEMPI \citep{KBKMR12c}.
Meridional cross-sections of $\meanB_\phi/\Beqz$ together with
magnetic field lines of $\meanBB_{\rm pol}$ are shown in
\Fig{pmer_bphi_256x1024P_qp100c_Q1e3_a006c_pole_str}.
Note that a magnetic flux concentration develops near the surface at
latitudes between $70\degr$ and $76\degr$ for weak and strong
stratification, respectively.
Structure formation from NEMPI occurs in the top 5\% by radius,
and the flux concentration is most pronounced when $\rstar\leq1.01$.

\begin{figure}[t!]\begin{center}
\includegraphics[width=\columnwidth]{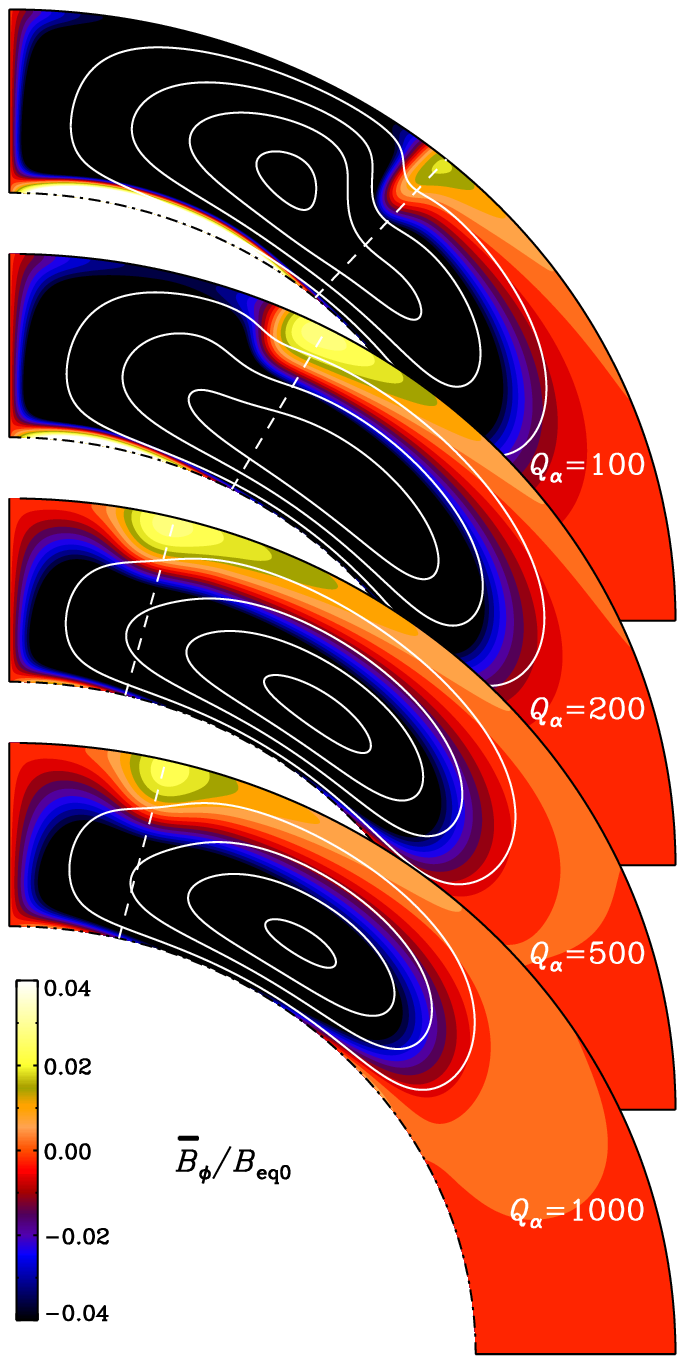}
\end{center}\caption[]{
Meridional cross-sections for different values of $Q_\alpha$,
for $\rstar=1.001$.
The dashed lines indicate the latitudes $49\degr$, $61.5\degr$, $75.6\degr$,
and $76.4\degr$.
}\label{pmer_bphi_256x1024P_qp100c_Qalp_a006c_pole}
\end{figure}

\begin{figure}[t!]\begin{center}
\includegraphics[width=\columnwidth]{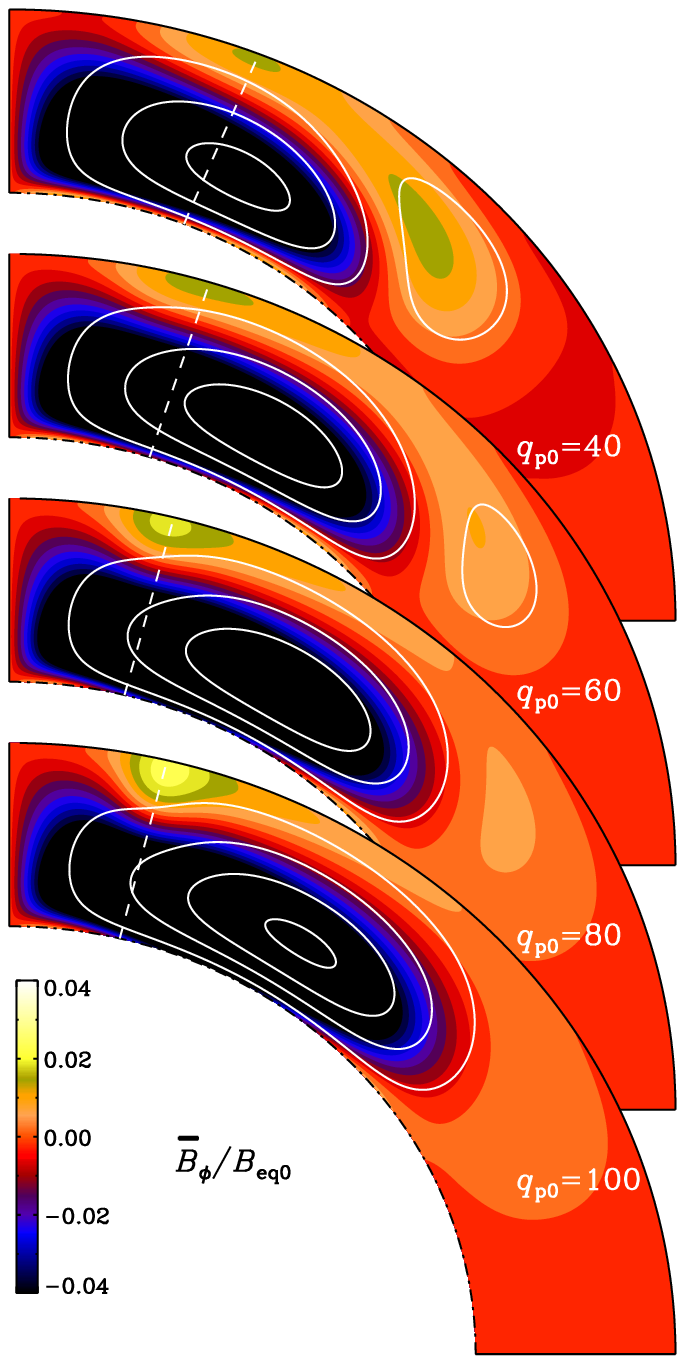}
\end{center}\caption[]{
Meridional cross-sections for different values of the parameter $\qpz$
in the range $40\leq\qpz\leq100$ for $Q_\alpha=10^3$.
The dashed lines indicate the latitudes $68\degr$, $72.5\degr$, $75.7\degr$,
and $76.3\degr$.
}\label{pmer_bphi_256x1024P_qp100c_Q1e3_a006c_pole_qp0b}
\end{figure}

Next, if we increase the magnetic field strength by making $Q_\alpha$
smaller, we see that the magnetic flux concentrations move toward lower
latitudes down to about $49\degr$ for $Q_\alpha=100$; see
\Fig{pmer_bphi_256x1024P_qp100c_Qalp_a006c_pole}.
However, while this is potentially interesting for the Sun, where sunspots
are known to occur primarily at low latitudes,
the magnetic flux concentrations also
become weaker at the same time, making this feature less interesting
from an astrophysical point of view.
For comparison with the parameter $\betap=0.05$ in \Eq{qp-apr} we note
that $\beta_\alpha=Q_\alpha^{-1/2}$ takes the values 0.1, 0.07, 0.04,
and 0.03 for $Q_\alpha=100$, 200, 500, and 1000, respectively.
Thus, for these models the quenchings of the nondiffusive turbulence
effects in the momentum and induction equations are similar.

Also, if we decrease $\qpz$ to more realistic values, we expect
the magnetic flux concentrations to become weaker.
This is indeed borne out by the simulations; see
\Fig{pmer_bphi_256x1024P_qp100c_Q1e3_a006c_pole_qp0b},
where we show meridional cross-sections for $\qpz$
in the range $40\leq\qpz\leq100$ for $Q_\alpha=10^3$.
This corresponds to the range $0.32\leq\betastar\leq0.5$.

\begin{figure}[t!]\begin{center}
\includegraphics[width=\columnwidth]{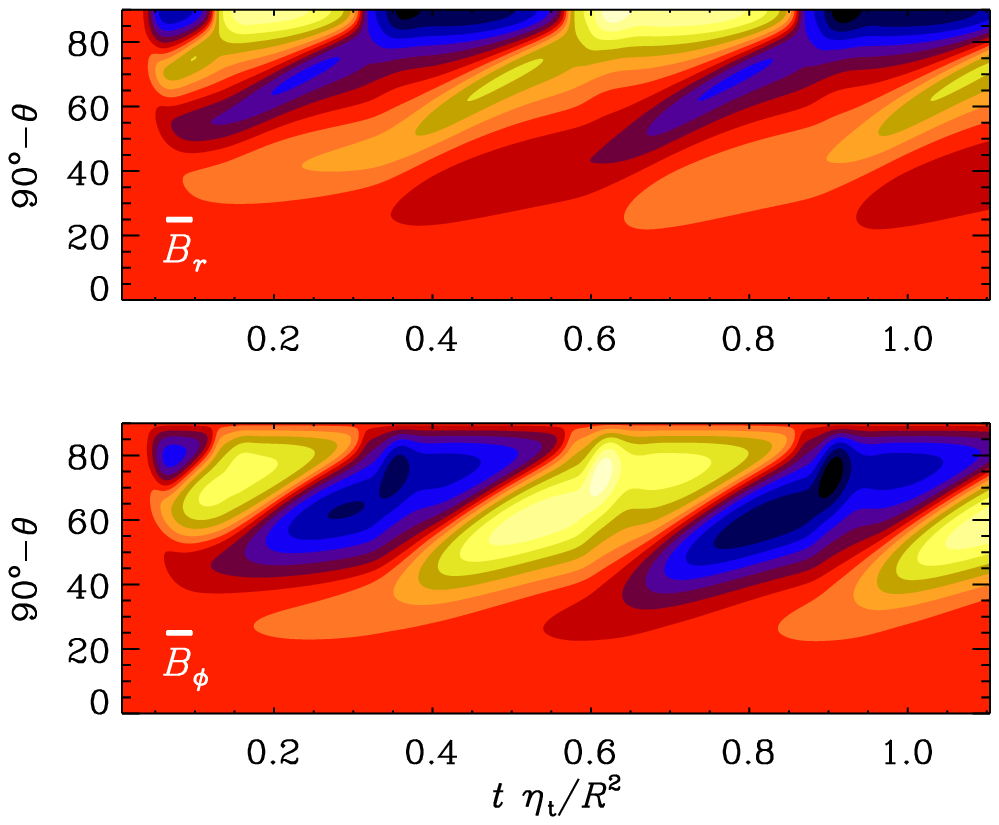}
\end{center}\caption[]{
Butterfly diagram of $\meanB_r$ (upper panel) and $\meanB_\phi$
(lower panel) for $Q_\alpha=10^4$, $\rstar=1.001$,
$\omega=11.3\,\etat/R^2$.
}\label{pbxmxy_256x1024P_qp100c_Q1e4_a006c_pole}
\end{figure}

\begin{figure}[t!]\begin{center}
\includegraphics[width=\columnwidth]{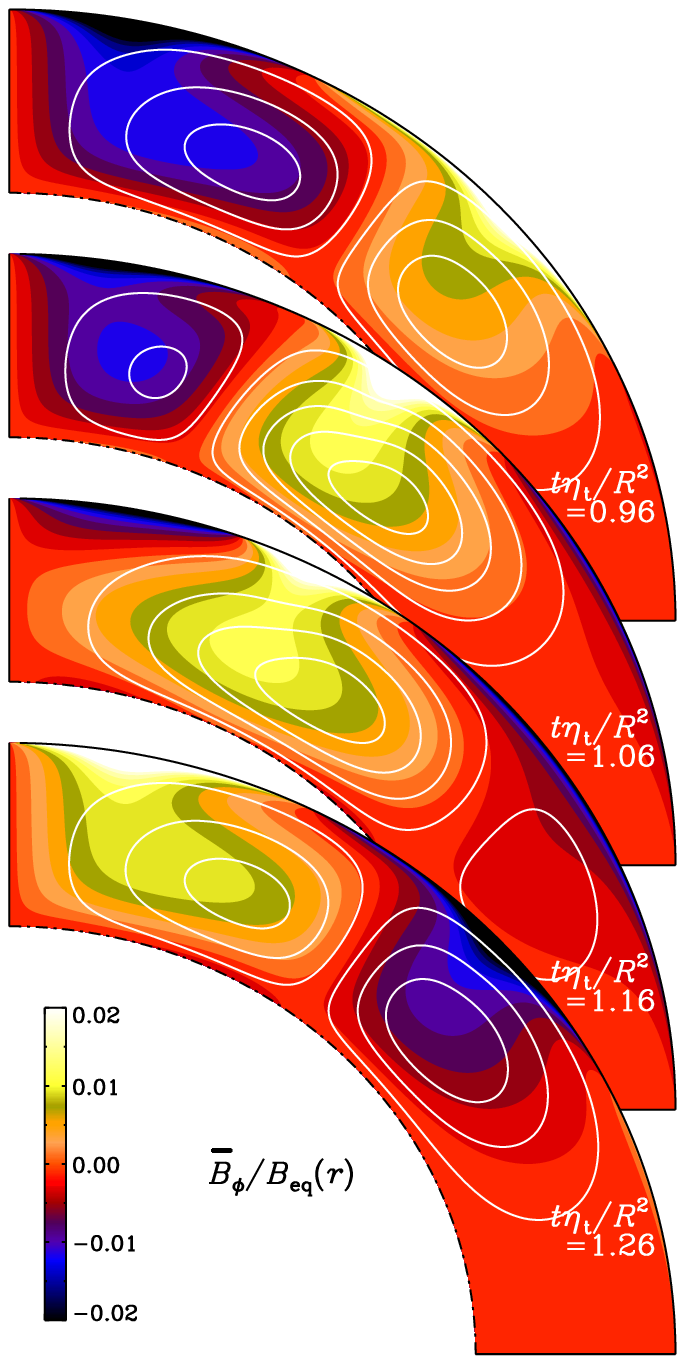}
\end{center}\caption[]{
Meridional cross-sections of $\meanBB/\Beqz$ at different times,
for $Q_\alpha=10^4$, $\rstar=1.001$.
The cycle frequency here is $\omega=11.3\etat/R^2$.
Furthermore, the toroidal field is normalized by the local equipartition
value, i.e., the colors indicate $\meanB_\phi/\Beq(r)$.
}\label{pmer_bphi_256x1024P_qp100c_Q1e4_a006c_pole_Beq}
\end{figure}

For weaker magnetic fields, i.e., for higher values of the quenching parameter
$Q_\alpha$, we find that NEMPI has a modifying effect on the dynamo
in that it can now become oscillatory.
A butterfly diagram of $\meanB_r$ and $\meanB_\phi$ is shown in
\Fig{pbxmxy_256x1024P_qp100c_Q1e4_a006c_pole}.
Meridional cross-sections of the magnetic field
at different times covering half a magnetic cycle are shown in
\Fig{pmer_bphi_256x1024P_qp100c_Q1e4_a006c_pole_Beq}.
It turns out that, at sufficiently weak magnetic field strengths, NEMPI
produces oscillatory solutions with poleward-migrating flux belts.
The reason for this is not understood very well, but it is reminiscent
of the poleward migration observed in the presence of weak rotation
\citep{Losada1}.
Had this migration been equatorward, it might have been tempting to
associate it with the equatorward migration of the sunspot belts in
the Sun.

\begin{figure}[t!]\begin{center}
\includegraphics[width=\columnwidth]{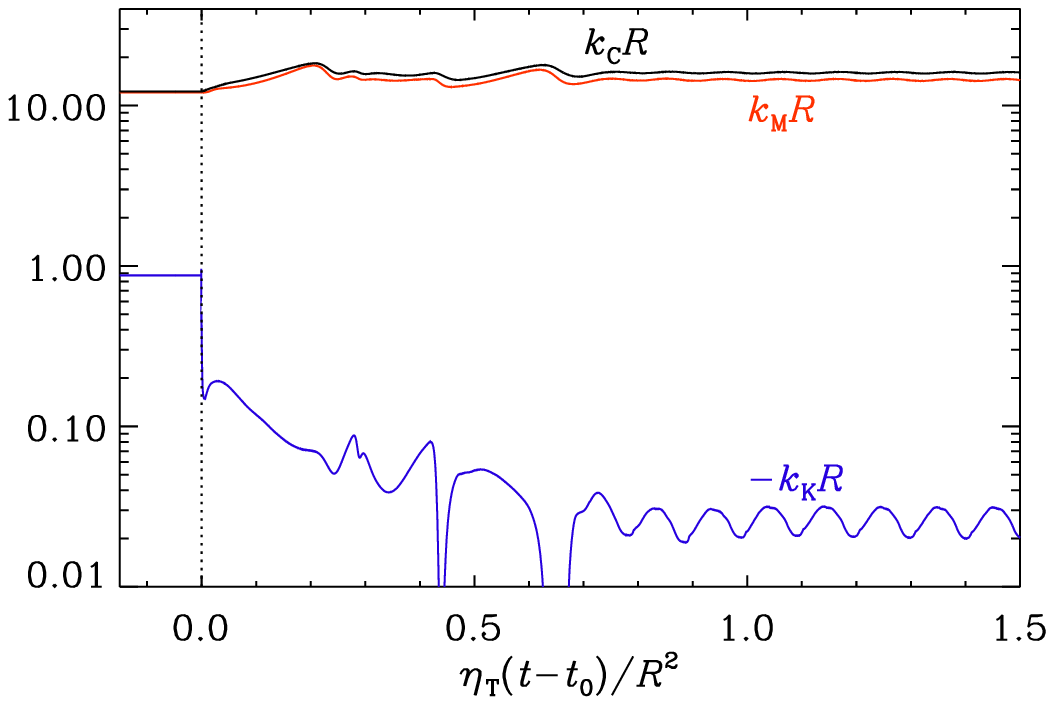}
\end{center}\caption[]{
The three inverse length scales $k_{\rm C}$, $k_{\rm M}$, and $k_{\rm K}$ as a function of time.
At time $t_0$, the value of $\qpz$ has been changed from 0 to 100.
}\label{phelicity}
\end{figure}

Finally, we discuss the change of kinetic, magnetic, and current
helicities due to NEMPI.
We do this by using a model that is close to our reference model with
$\rstar/R=1.001$ and $Q_\alpha=1000$, except that $\qpz=0$ in the beginning,
and then at time $t_0$ we change it to $\qpz=100$.
The two inverse length scales based on magnetic and current helicities,
\EQ
k_{\rm M}=\left({\int_V\meanAA\cdot\meanBB\,\dd V\over
\int_V\meanBB^2\,\dd V}\right)^{-1}\;\mbox{and}\quad
k_{\rm C}=\mu_0{\int_V\meanJJ\cdot\meanBB\,\dd V\over
\int_V\meanBB^2\,\dd V},
\EN
increase by 25\%, while the inverse length scale based on the kinetic helicity,
\EQ
k_{\rm K}={\int_V\meanWW\cdot\meanUU\,\dd V\over
\int_V\meanUU^2\,\dd V},
\EN
drops to very low values after introducing NEMPI, see e.g.\ \Fig{phelicity}.
Here, $\meanWW=\nab\times\meanUU$ is the mean vorticity.
This behavior of $k_{\rm K}$ is surprising, but it seems to be associated with an increase
in kinetic energy.
The reason for the increase in the two inverse magnetic length scales, on the
other hand, might be understandable as the consequence of increasing gradients
associated with the resulting flux concentrations.

\section{Conclusions}

The present investigations have shown that NEMPI can occur
in conjunction with the dynamo; that is,
both instabilities can work at the same time and
can even modify each other. It was already clear
from earlier work that NEMPI can only work in a
limited range of magnetic field strengths. We
therefore adopted a simple $\alpha$ quenching
prescription to arrange the field strength to be
in the desired range. Furthermore, unlike much of
the earlier work on NEMPI, we used an adiabatic
stratification here instead of an isothermal one;
see \cite{BKR10} and \cite{Kapy12} for earlier
examples with adiabatic stratification in
Cartesian geometry. An adiabatic stratification
implies that the pressure scale height is no
longer constant and now much shorter in the upper
layers than in the bulk of the domain. This
favors the appearance of NEMPI in the upper
layers, because the growth rate is inversely
proportional to the pressure scale height.

There are two lines of future extensions of the present model.
On the one hand, it is important to study the interplay
between NEMPI and the dynamo instability in more detail.
This is best done in the framework of a local Cartesian model, which is
more easily amenable to analytic treatment.
Another important extension would be to include differential rotation.
At the level of a dynamically self-consistent model, where the flow speed
is a solution of the momentum equation, differential rotation is best
implemented by including the $\Lambda$ effect \citep{Rue80,Rue89}.
This is a parameterization of the Reynolds stress that is in some ways
analogous to the parameterization of the electromotive force via the
$\alpha$ effect.

Mean-field models with both $\alpha$ and $\Lambda$ effects have been
considered before \citep{BMT92,Rem06}, so the main difference would be the
additional parameterization of magnetic effects in the Reynolds
stress that gives rise to NEMPI.
In both cases, our models would be amenable to verification using DNS
by driving turbulence through a helical forcing function.
In the case of a spherical shell, this can easily be done in wedge
geometry where the polar regions are excluded.
In that case the mean-field dynamo solutions are oscillatory
with equatorward migration \citep{Mitra}.
At an earlier phase of the present investigations we studied NEMPI
in the corresponding mean-field models and found that NEMPI can
reverse the propagation of the dynamo wave from equatorward to poleward.
However, owing to time dependence, the effects of NEMPI are then harder
to study, which is why we have refrained from studying such models in
further detail.

In the case of a Cartesian domain, helically forced DNS with an open
upper layer have been considered by \cite{WB10}.
In this model, plasmoid ejections can occur and provide a more
natural boundary.
A more physical alternative is to use only nonhelical forcing, but to
include rotation to produce helicity in conjunction with the stratification.
Such models have recently been considered by \cite{Losada2}, who found
that NEMPI begins to be suppressed by rotation at Coriolis numbers
somewhat below those where $\alpha^2$-type dynamo action sets in.
Furthermore, there is now evidence that the combined action of NEMPI
and the dynamo instability has a lower threshold than the dynamo alone.
Those models provide an ideal setup for future studies of the interaction
between both instabilities.

\begin{acknowledgements}
This work was supported in part by the European Research Council under the
AstroDyn Research Project No.\ 227952,
by the National Science Foundation under Grant No.\ NSF PHY05-51164 (AB),
by EU COST Action MP0806,
by the European Research Council under the Atmospheric Research Project No.\
227915, and by a grant from the Government of the Russian Federation under
contract No. 11.G34.31.0048 (NK, IR).
We acknowledge the allocation of computing resources provided by the
Swedish National Allocations Committee at the Center for
Parallel Computers at the Royal Institute of Technology in
Stockholm and the Nordic Supercomputer Center in Reykjavik.
\end{acknowledgements}


\end{document}